\documentclass{ecai} 
\usepackage{latexsym}
\usepackage{amssymb}
\usepackage{amsmath}
\usepackage{amsthm}
\usepackage{booktabs}
\usepackage{enumitem}
\usepackage{graphicx}
\usepackage{color}
\usepackage{amsfonts}
\usepackage{algorithm}
\usepackage{array}
\usepackage{makecell}
\usepackage{multirow}  %for the tables
\usepackage{xurl} %for breaking long urls
\usepackage{bbm} %for mathbbm
\usepackage{subfig} %for figures
\usepackage[colorlinks=false]{hyperref} %for \url
\usepackage{xcolor}
\hypersetup{
    colorlinks,
    linkcolor={red!50!black},
    citecolor={black!50!black},
    urlcolor={black!80!black}
}

\newtheorem{theorem}{Theorem}

\newtheorem{definition}{Definition}
\newtheorem{example}[theorem]{Example}
\theoremstyle{remark}

\usepackage{romannum}
\AtBeginDocument{\pagenumbering{arabic}}

\newcommand{\BibTeX}{B\kern-.05em{\sc i\kern-.025em b}\kern-.08em\TeX}
\usepackage[noend]{algpseudocode}
\algnewcommand{\Inputs}[1]{%
  \State \textbf{Input:}
  \Statex \hspace*{\algorithmicindent}\parbox[t]{.8\linewidth}{\raggedright #1}
}
\algnewcommand{\Initialize}[1]{%
  \State \textbf{Initialize:}
  \Statex \hspace*{\algorithmicindent}\parbox[t]{.8\linewidth}{\raggedright #1}
}
\algnewcommand{\Outputs}[1]{%
  \State \textbf{Output:}
  \Statex \hspace*{\algorithmicindent}\parbox[t]{.8\linewidth}{\raggedright #1}
}

\begin{document}

%%%%%%%%%%%%%%%%%%%%%%%%%%%%%%%%%%%%%%%%%%%%%%%%%%%%%%%%%%%%%%%%%%%%%%%%

\begin{frontmatter}

%%% Use this command to specify your submission number.
%%% In doubleblind mode, it will be printed on the first page.

%\paperid{265} 

%%% Use this command to specify the title of your paper.

\title{Interactive and Iterative Peer Assessment}

%%% Use this combinations of commands to specify all authors of your 
%%% paper. Use \fnms{} and \snm{} to indicate everyone's first names 
%%% and surname. This will help the publisher with indexing the 
%%% proceedings. Please use a reasonable approximation in case your 
%%% name does not neatly split into "first names" and "surname".
%%% Specifying your ORCID digital identifier is optional. 
%%% Use the \thanks{} command to indicate one or more corresponding 
%%% authors and their email address(es). If so desired, you can specify
%%% author contributions using the \footnote{} command.

\author[A]{\fnms{Lihi}~\snm{Dery}\orcid{0000-0002-8710-3349}\thanks{Email: lihid@ariel.ac.il}}

\address[A]{Ariel University, Israel}

%%% Use this environment to include an abstract of your paper.

\begin{abstract}
Iterative peer grading activities may keep students engaged during in-class project presentations. 
Effective methods for collecting and aggregating peer assessment data are essential. Students tend to grade projects favorably. So, while asking students for numeric grades is a common approach, it often leads to inflated grades across all projects, resulting in numerous ties for the top grades. Additionally, students may strategically assign lower grades to others' projects so that their projects will shine. Alternatively, requesting students to rank all projects from best to worst presents challenges due to limitations in human cognitive capacity. To address these issues, we propose a novel peer grading model consisting of (a) an algorithm designed to elicit student evaluations and (b) a median-based voting protocol for aggregating grades to a single ranked order that reduces ties. An application based on our model was deployed and tested in a university course, demonstrating fewer ties between alternatives and a significant decrease in students' cognitive and communication burdens.
\end{abstract}

\end{frontmatter}

%%%%%%%%%%%%%%%%%%%%%%%%%%%%%%%%%%%%%%%%%%%%%%%%%%%%%%%%%%%%%%%%%%%%%%%%

\section{Introduction}\label{sec:intro}

%what it is
Peer assessment is a structured procedure where individuals in a group evaluate the performance of their peers who share a similar status within that group~\citep{topping1998peer}.
This study concentrates on peer assessment in an academic classroom~\citep{topping2009peer}, and specifically on the quantitative peer assessment (or peer grading) of oral class presentations~\citep{Grez_Valcke_Berings_2010,Yu_Liu_Liu_2023}. 
In this educational setting, students present their projects during class and have the opportunity to evaluate their peers' work and receive assessments themselves. 
During the assessment activity, students are organized into distinct sessions; in each session, students take turns presenting their work. Peer assessment is done with the understanding that students' performance as assessors and presenters contributes to their final grades.

\begin{figure}[h] 
\centering
\includegraphics[width=7.5cm]{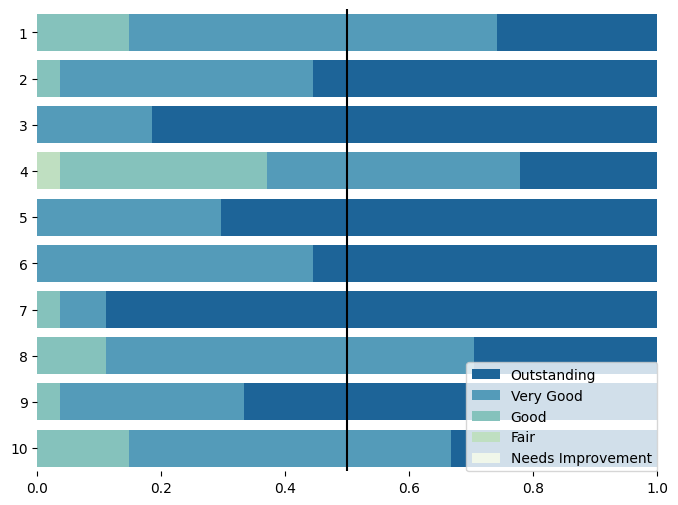}
\caption{The grade distribution of ten projects in a university course. The y-axis shows the project numbers, and the x-axis shows the cumulative percentage of students who gave these projects a particular grade, as specified in the legend. The vertical black line indicates the median score.}
\label{fig:session a}
\vspace{0.3cm}
\end{figure}

%gap
Even though some studies have described peer assessment systems, and there exists a survey of tools (albeit a nearly 15-year-old one)~\citep{luxton2009systematic}, little attention has been paid to the methodologies used for collecting and aggregating peer assessment data.
These methodologies are the focus of this research. They are needed since several challenges emerge when students are requested to assign scores (ratings) to projects.
The first two may also arise in other rating scenarios, and the latter is specific to peer assessment settings:
%\begin{itemize} 

     \textbf{(a) Calibration and standardization.} Rating on a wide range numerical scale (e.g., $0-100$) is not always meaningful for students, who might have difficulty in choosing between several close scores (e.g., between $86$ and $87$). To solve this issue, ratings are typically conducted on a 5-point or 7-point Likert scale. 
    The scale alone is insufficient, as there is no standard agreement on the meaning of the numbers on these scales. One student may perceive a rating of $5$ as another perceives a $4$. To reach a common understanding, some calibration or elimination of bias must be performed (e.g., ~\citep{piech2013tuned,song2016toward, rico2018statistical}).    
    Thus, in this research, a grading language~\citep{balinski2011majority} or scoring rubric~\citep{arter2001scoring} is employed on top of the scale. Specifically, ``Outstanding" is mapped to the grade $5$, and ``Very Good" to $4$.

     \textbf{(b) Strategization.} Students have been reported to employ subjective underscoring as a strategy, particularly under rivalry and competitive circumstances~\citep{baruah2018should,meijer2020unfolding}. Consider, for example, three student projects: $c_1$, $c_2$, and $c_3$. An assessor who favors $c_1$ will assign it the highest possible score, but to ensure that $c_1$ ends up in the first place, the assessor might also assign $c_2$ and $c_3$ the lowest possible score. A partial solution adopted in this research is to compute the median rating instead of the average rating, as it is not hard to see that the median is more immune to this sort of manipulation.        
    
    \textbf{(c) Generosity.} Students struggle to be critical towards their peers~\citep{lindblom2006self} and often assign generous grades, while teachers are more strict~\citep{nejad2019assessment,panadero2013impact}. This student tendency to over-mark has been seen to occur when students have no anonymity since they do not want to be seen as penalizing their peers~\citep{vickerman2009student}. We found this also occurs in anonymous settings; Figure~\ref{fig:session a} illustrates the grades that $27$ students gave to $10$ projects their peers presented in class.
    The median score for all projects on a rubric of five was either ``Outstanding'' or ``Very Good''; none of the projects received a median grade of ``Good'' or lower. 
    This is not an irregularity. Peer grading data from ten different class sessions all exhibited similar performances 
    (see Appendix
    \footnote{\url{https://doi.org/10.17605/OSF.IO/K4YMA}}
    ).  
    This does not mean that students think all projects are equally outstanding. As we show later, when prompted, students are indeed able to state which projects they prefer over others. %However, simply requiring students to assign grades may not be sufficient.
    
To address these challenges, some peer assessment systems have suggested collecting ordinal preferences~\citep{raman2014methods} or assigning distinct scores to each project ( effectively translating into ranking the alternatives).
However, experts and non-experts alike are limited in the number of preferences they can meaningfully order ---
sometimes, they simply do not hold a rank over the alternatives (see, e.g.
Chapter 15 in ~\citet{balinski2011majority}), and oftentimes they 
find it hard to rank more than seven items from best to worst~\citep{miller1956magical}.
Pairwise comparison queries such as ``Do you prefer this alternative over that alternative?'' may assist people in forming a ranking~\citep{dery2014reaching,lu2011learning}. However, this can be perceived as a tiresome task for voters, as the required number of comparison queries is large (e.g., 45 queries for ten projects, 105 queries for 15 projects). 

In summary, limiting students to rating projects results in the issues detailed above, while limiting students to ranking alternatives requires a cognitive and communication effort that may burden them.

\textbf{Contributions:}
In this paper, 
we study peer assessment from an algorithmic perspective. 
We integrate both cardinal preferences, expressed through ratings or grades, and ordinal preferences, expressed through pairwise comparisons.
Our contribution is a novel hybrid peer assessment model, $R2R$, comprising (a) an algorithm that elicits student ratings and conducts pairwise comparison queries as needed and (b) a method for combining grades to create one overall ranking. The class instructor can then utilize this ranking to evaluate the projects and select the top projects in the class. 

\section{Related work}\label{sec:lit}

The importance of peer assessment transcends mere grading by students and is an integral component of the learning process, commonly referred to as ``assessment for learning'' rather than ``assessment of learning''~\citep{tuomainen2023supportive}.
There exists a large body of literature on the  benefits of peer assessment 
(e.g., ~\citep{adesina2023managing, chang2020integration,chien2020effects,double2020impact,li2020does,li2021effect,reddy2021student,yan2022effects}).
%(e.g., ~\citep{adesina2023managing, chang2020integration,chien2020effects,double2020impact,li2020does,li2021effect,reddy2021student,seifert2019online,yan2022effects}).

Ranking-based systems have better reliability~\citep{shah2013case}, and assessors using these systems have been found to be 10\% more reliable than rating-based systems~\citep{song2017exploratory}.
Some systems allocate varying weights to reviewers based on their credibility (e.g., \citep{tao2018domain}). 
Others implement controls on who reviews whom or incorporate AI-based methods \citep{darvishi2022incorporating}.
\citet{walsh2014peerrank} proposes a peer-rank rule in which the grade a user assigns to another user is weighted by the assigning user's own grade. The user's grade is the weighted average of the grades of the users evaluating that user. This approach shares similarities with the PageRank algorithm \citep{page1998pagerank}.
However, a prevailing perspective in peer assessment is that all assessors (students) should be considered equally credible.
Consequently, most peer assessment systems aggregate ratings using either the average or the median rating (e.g.,~\citep{pare2008peering, purchase2018peer, wind2018peer}).

Simply computing the average rating is not a good solution, as it is easily manipulable~\citep{stelmakh2021catch}. This is especially true in peer assessment systems, as the users of the system are also the candidates (or friends of the candidates) and may attempt to strategize, as explained in the previous section.  
While averaging is commonly used in sports performance evaluations, entities like the Olympic Committee and international federations often eliminate the highest and lowest ratings received by a candidate. In the specific context of sports performance evaluation,~\citet{osorio2020performance} introduces a bias correction procedure that addresses deviations from the mean score and individual users' grading patterns. 
However, these methods do not perform well in the context of peer assessment, because there is typically little variance in peer assessment ratings, and the evaluators are often students without an extensive voting history.

Computing the median rating offers a less manipulable solution that works in peer assessment context, however, it can lead to numerous ties between alternatives. 
The Majority judgment voting protocol~\citep{balinski2011majority} addresses the issue of tie-breaking the median, with recent methods proposed for this purpose by~\citet{fabre2021tie}. Nevertheless, these methods still result in ties when applied to student peer rating grades. In this research, the median is initially computed, but to overcome the limitations associated with ties, rankings are also integrated into the assessment process.

% closely related fields
Peer assessment has extended to applications that involve a large number of items, where obtaining evaluations from each user on every item becomes impractical. These applications include massive online courses (MOOCs)~\citep{kulkarni2013peer,piech2013tuned}, freelancing platforms~\citep{kotturi2020hirepeer}, and academic conference reviews(~\citep{stelmakh2021catch}). However, the primary algorithmic focus of these applications lies in determining who reviews what and in aggregating incomplete sets of reviews. In our case, we assume that all students evaluate all projects, resulting in complete preference sets.
% Peer assessment is a sub-field in peer prediction, or preference elicitation without verification~\citep{miller2005eliciting}. 
% In the latter, one challenge is incentivizing users to provide their truthful preferences~\citep{burrell2023measurement}.

Herein, we assume that all students are equal and all students rate all projects.  
Possible bias in evaluations is mitigated by employing  a common grading language as done by~\citet{balinski2011majority} and by using
ordinal as well as cardinal evaluations.

% \subsection{bias}
% %another thought
% It is a barrier against miscalibiration in ratings (see another method for handling this: \citep{wang2019your}), since the ratings are turned into a ranking. 

\subsection{Median-based voting rules}\label{subsec:tie_breaks}
We provide a short summary of state-of-the-art median-based voting methods, as our model builds upon them.

Let there be $n$ voters (students) $V=\{v_1, v_2,\ldots,v_n\}$ and $m$ candidates (projects) 
$C=\{c_1, c_2,\ldots,c_m\}$.
Let $s_i^j$ denote the score assigned by voter $v_i$ to candidate $c_j$. 
The score is assigned from a predefined domain of discrete values 
$D = \{d_{min} \dots d_{max}\}$ where $d_{min}$ and $d_{max}$ are the lowest and highest values respectively $(d_{min}\leq s_i^j \leq d_{max})$.
The ordered set of all scores assigned to candidate $c_j$ is $S^j$.
The median of $S^j$  is the score in the middle-most position when $n$ is odd, and the score in position $n/2$ when $n$ is even.
As illustrated in Figure \ref{fig:session a}, the median score of most projects is ``Outstanding'' or ``Very Good'' (these scores translate into the numerical scores of ``5'' and ``4''). Therefore, projects cannot be ranked solely according to the median, and a tie-breaking mechanism is required.

A few mechanisms exist for tie-breaking the median between projects. They are all based on the same idea: first, compute the projects's median score. Then add a quantity based on what~\citet{fabre2021tie} terms as proponents ($p$) and opponents($q$). 
The proponent $p$ is the share of scores higher than the median $\alpha$:
    $p_j = \sum\limits_{i|s_i^j>\alpha}s_i^j$.  
    The opponent $q$ is the share of scores lower than the median $\alpha$:
    $q_j = \sum\limits_{i|s_i^j<\alpha}s_i^j$.

The mechanisms differ in their use of $p$ and $q$:  
a candidate's score is the sum of its median grade $\alpha$ and a tie-breaking rule. \citet{fabre2021tie} defines three rules: Typical judgment, Central judgment, and Usual judgment. 

\begin{definition}[Typical judgement]
  Typical judgment is based on the difference between non-median groups:
  $c^{\Delta} = \alpha + p - q$.
\end{definition}

\begin{definition}[Central judgement]
    Central judgment is based on the relative share of proponents:
    $c^{\sigma} = \alpha + 0.5\cdot\frac{p-q}{p+q}$. When $p=q=0$ we set $c^{\sigma} = \alpha$.
\end{definition}

\begin{definition}[Usual judgement]
    Usual judgment is based on the normalized difference between non-median groups: $ c^{v} = \alpha +  0.5\cdot\frac{p-q}{1-p-q}$.
\end{definition}

\citet{fabre2021tie} shows that majority judgment~\citep{balinski2011majority} can also be defined using just $\alpha, p$ and $q$ :
\begin{definition}[Majority judgement]
     $c^{mj} = \alpha + \mathbbm{1}_{p>q} - \mathbbm{1}_{p\leq q}$.
     \footnote{$\mathbbm{1}_{f(x)}$ is an indicator function of $f(x)$. E.g., if $p>q$ then $\mathbbm{1}_{p>q} = 1$}
     If ties remain, the median of the tied candidates is dropped, and then $mj$ is recomputed for the tied candidates. This procedure is repeated until all ties are resolved. Note that for ties \citet{fabre2021tie} suggest an alternative method that results in the same output. 
\end{definition}

\begin{example}
    Let us consider the set of scores assigned to candidate $c_j$ by ten voters:
    \[
    S^j = \{ 5, 5, 5, 4, 4, 4, 3, 3, 3, 2 \}
    \]

    The calculated median for these scores is $\alpha = 4$. Among the voters, three assigned scores above the median, while four assigned scores below the median. This leads us to define the proponent and opponent fractions as $p=\frac{3}{10}$ and $q=\frac{4}{10}$, respectively.

    For the candidate $c_j$, the scores according to Typical, Central, Usual, and Majority judgments can be determined as follows:

    \[
    c_j^{\Delta} = 4 + \frac{3}{10} - \frac{4}{10} = 3.9
    \]

    \[
    c_j^{\sigma} = 4 + 0.5\cdot\frac{-0.1}{0.7} \approx 3.92
    \]

    \[
    c_j^{v} = 4 + 0.5\cdot\frac{-0.1}{1.1} \approx 3.83
    \]

    \[
    c_j^{mj} = 4 - 1 = 3
    \]
\end{example}

Project scores can be computed using the rules outlined above. A ranking of the projects can be obtained by simply ordering them according to their scores.

\section{The model}
%Voters are usually asked to either rank or rate alternatives. However, we claim that reducing their task to just this or the other conceals important information about their full preferences. 
Our proposed model has two key components: $R2R$ Elicitation and $R2R$ Aggregation. Together, these components create a hybrid model that incorporates both rankings and ratings.

\subsection{R2R elicitation}
The $R2R$ Elicitation phase aims to gather voter preferences while minimizing cognitive and communication overload. Voters are requested to provide ratings for each candidate, with pairwise comparison queries employed only when necessary. We introduce the concept of a \textit{Ranked Rating Set} to capture both scores and rankings.

\begin{definition}[Rating Set]
A rating set $S_i$ contains the scores that voter $v_i$ assigned to candidates $c_1 \dots c_m$.
\end{definition}

\begin{definition}[Preference profile]
A preference profile $\pi_i \in \Pi$ is a tuple $<c,p>$ containing candidates ($c$) and their ranked position ($p$) according to $v_i$. 
We write $c_i \succ c_j$ when $c_i$ is preferred over $c_j$. For $k<l$ the preference profile is thus:  $\{\ldots (c_i,p_k),(c_j,p_l) \ldots \}$. 
\end{definition}

When $v_i$ assigns a different score to each candidate, then the preference profile can be constructed directly from these scores. For example, when 
$ S_i = \{s_i^1 = 4,  s_i^2 =5,  s_i^3 =3 \}$ then the preference profile $\pi_i$ is: 
$\{(c_2,1),(c_1,2),(c_3,3) \}$
or equivalently
$c_2 \succ c_1 \succ c_3$.

It is usually assumed that we can elicit the voter's rating set or the voter's preference profile. Here, we deviate from the standard literature and define a ranked rating set. A ranked rating set contains attributes of both a rating set and a preference profile, thus allowing us to save more information about voters' preferences. 

\begin{definition}[Ranked Rating Set]
A ranked rating set $\psi \in \Psi$ is a tuple $<c,s,p>$ containing candidates ($c$), their scores ($s$) and their position ($p$). As in the ranking set, the position is determined by the ranking. 
The rating set of voter $v_i$ is denoted $\psi_i$. 
\end{definition}

If $v_i$ assigns the same score to two or more candidates, a pairwise comparison query $q(v_i,c_j,c_k) \in Q$ is executed. 
We assume that a voter can always respond to a query, i.e.,
a pairwise comparison query $q(v_i,c_j,c_k)$ has only two possible responses: $c_j \succ c_k$ or $c_k \succ c_j$.

We present a method, Rating to Ranking, $R2R$, which builds a ranked rating set while eliciting the needed information from the voters. 
A pseudo-code is provided in Algorithm \ref{alg:R2R} followed by a detailed description and a running example.

\begin{algorithm}[h]
	\caption{Rating to Ranking (R2R) elicitation}
	\label{alg:R2R}
	\begin{algorithmic}[1]
		\Inputs{a set of candidates: $c_1, c_2 \ldots, c_m$ \\
	        	a set of voters: $v_1, v_2 \ldots, v_n$ \\
	        	a score domain $D = \{d_{min} \dots d_{max}\}$ \\
    	        }
        \For{$v_i \in V$}  %iterate on voters 
        \State $\psi_i \in \Psi \gets []$ %begin with an empty rating set
            \For{$c_j \in C$} %iterate on cands
	            \State  voter declares $s_i^j \in D$   %voter assigns score to cand
	            %%%\State $s_i \gets insert(c_j, d_i^j)$
	            \State $count \gets count(\psi_i, s_i^j) $ %number of candidates with score value $d$
    	        \If {$count = 0$}
    	            \State $insert(\psi_i, c_j)$
    	        \Else %if there are more than one cands with this score
    	            \State $p \gets lastIndex(\psi_i, s_i^j)$ 
        	        \While{$(count \geq 1)$}  
                        \State $c_p \gets candAtIndex(\psi_i, p)$
                        \State execute query $q(v_i,c_j,c_p)$  %issue a query
                        \If {$c_j \succ c_p$}
                            \State {$p--$}
                            \State {$count--$}
                        \Else 
                            \State exit while loop
                        \EndIf
                    \EndWhile
                    \State $insert(\psi_i, c_j, p)$
                \EndIf
           \EndFor
        \EndFor
	    \Outputs{$\psi_i$ - a ranked rating set for each voter $v_i$}
	\end{algorithmic}
\end{algorithm}

The algorithm receives as input a set of $m$ alternatives, $n$ voters, and a score domain $D$.
Each voter is sequentially asked to provide scores for all candidates (lines 2-5). Note that this process can also be done the other way around: for each candidate, all voters must submit their scores. In either case, voter $v_i$ assigns a score $s_i^j$ to candidate $c_j$ (line 5). 
Subsequently, the algorithm counts how many times this particular score ($s_i^j$) has been previously submitted by the voter (line 6).
If the voter has not submitted this score before, it is inserted to $\psi$ (lines 7-8) while ensuring its ordered placement (lines 7-8), utilizing the $insert$ function.
If the score $s_i^j$ already exists within $\psi$, the algorithm retrieves the index of the last position where it is found (line 10) and identifies the candidate $c_p$ at that position (line 12). A pairwise query is then executed to determine the relative ranking between the two candidates (line 13). 
If $c_j$ is ranked higher than $c_p$ ($c_j \succ c_p$) the algorithm proceeds to the next position and repeats the query process (lines 14-16, followed by line 11). Otherwise, the algorithm proceeds to insert $c_j$ at the identified position $p$ (lines 19). 
Finally, the output is a ranked rating set (line 20). 

It is important to note that this pairwise comparison approach prevents the elicitation of Condorcet cycles.

\begin{example}
Consider one voter $v_1$ and four candidates: $c_1, c_2, c_3, c_4$. Scores are given in domain $D = {1,2 \ldots 5}$.
Assume that 
the voter's rating set is: $S_1 = \{s_1^1 = 5, s_1^2=4, s_1^3=5, s_1^4 = 5\}$
and the voter's preference profile is: $\pi_1 = \{  c_3 \succ c_4 \succ c_1 \succ c_2 \}$.
The rating set and the preference profile are initially unknown. The goal is to determine the ranked rating set. 

At first, $v_1$ declares her score for $c_1$: $s_1^1 = 5$. Since this is the first score declared, the ranked rating set is updated to $\psi_1 = \{(c_1,5,0) \}$.
Then, the next score is declared: $s_1^2 = 4$. Since $\psi$ does not contain this score, $\psi_1$ is updated without any queries: $\psi_1 = \{(c_1,5,0), (c_2,4,1) \}$. The next score declared is $s_1^3 = 5$. Since this score is already in $\psi_1$, a query is issued: $q(v_1,c_1,c_3)$. The voter responds by stating: $c_3 \succ c_1$. 
Thus $c_3$ is now added and $\psi_1$ positions of candidates is updated: $\psi_1 = \{\mathbf{(c_3,5,0)}, (c_1,5,\mathbf{1}), (c_2,4,\mathbf{2}) \}$ 
(changes to $\psi_1$ are marked in bold). 
Lastly, $v_1$ declares her score for $c_4$: $s_1^4 = 5$. This causes a query to be issued: $q(v_1,c_1,c_4)$. The voter responds with: $c_4 \succ c_1$, so yet another query is issued: $q(v_1,c_3,c_4)$. The voter responds with: $c_3 \succ c_4$ and $\psi_1$ is finalized:
$\psi_1 = \{(c_3,5,0), \mathbf{(c_4,5,1)}, (c_1,5,\mathbf{2}), (c_2,4,\mathbf{3}) \}$ (again, changes to $\psi_1$ are marked in bold).
\end{example}

%The algorithm ensures that a minimal number of queries is issued. We leave the proof to an extended version of this paper.  

\subsection{R2R aggregation}\label{subsec:agg}
%Median plus Copeland or Borda

The $R2R$ Aggregation phase combines voter preferences obtained through the $R2R$ Elicitation phase. Initially, the median $\alpha$ is calculated from the scores in the Ranked Rating Sets $\Psi$. 
The projects are grouped into buckets according to their medians. For all projects that share the same median, some voting rule is used to define their order. We herein employ two simple, well-known methods. 
Let $N(c_j,c_k)$ denote the number of voters who prefer $c_j \succ c_k$.
In Borda voting, the candidate's score is based on its ranking in the voters' preference order. Formally: 
\begin{definition}[$Borda$ voting rule]
The score of candidate $c_j$ is 
the the total number of voters who prefer $c_j$ over each other candidate $c_k$: 
$c_{j}^{borda} =
\sum_{k=1}^m,_{\forall k \neq j} N(c_j,c_k) $
\end{definition}

\citet{copeland1951reasonable} suggested a voting system based on pairwise comparisons. The method finds a Condorcet winner when such a winner exists. 
Each candidate is compared to every other candidate. The candidate obtains one point when it is preferred over another candidate by the majority of the voters. Adding up the points results is the Copeland score of each candidate. 
A family of Copeland methods, $Copeland^{\alpha}$, was suggested by~\citet{faliszewski2009llull}. 
The difference is in tie-breaking the alternatives. The fraction of points each candidate receives in case of a tie between two candidates can be set to any rational number: $0 \leq \alpha \leq 1$. The Copeland score of a candidate is thus the number of points it obtained plus the number of ties times $\alpha$.
\footnote{The Copeland $\alpha$ differs from the median $\alpha$. To maintain consistency with prior research and to aid readers familiar with the literature, we use the same symbol, $\alpha$, for both quantities and specify which $\alpha$ is being referenced.} 
Formally:

\begin{definition}[$Copeland^{\alpha}$ voting rule]
The score of candidate $c_i$ is its sum of victories in pairwise comparisons:

$c_{j}^{copeland^{\alpha}} =
\sum_{k=1}^m,_{\forall k \neq j}[ N(c_j,c_k) > n/2] + 
\alpha \cdot \sum_{k=1}^m,_{\forall k \neq j}[ N(c_j,c_k) \equiv  N(c_k,c_j)]$    

\end{definition}

$R2R\_b$ and $R2R\_c$ represent instances of $R2R$ that utilize Borda and Copeland voting, respectively, in the aggregation phase. 
While there are numerous ranking-based voting methods, we chose to begin with these methods since they are well-known and relatively easy to explain. Nonetheless, other voting methods are also applicable.

\section{User Study - R2R System}\label{sec:user_study}
We implemented a peer-rating system according to the $R2R$ model.
\footnote{The code is readily available at \url{https://github.com/nlihin/R2R}. The online system is at 
\url{https://rate2rank-0d561bf6674a.herokuapp.com/}}.
Interested parties can utilize the system upon request.

\begin{figure}[h] 
\centering
\includegraphics[scale = 0.22]{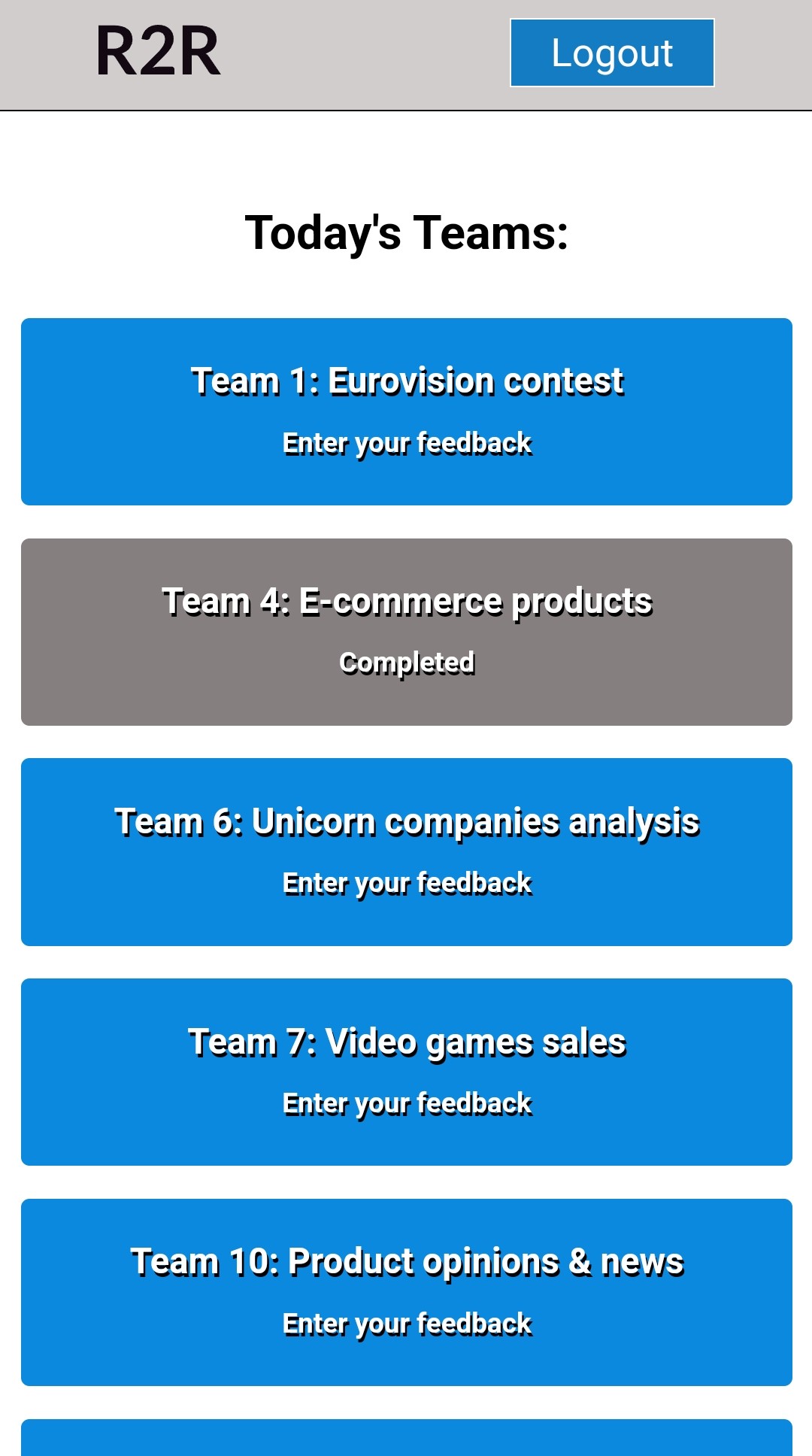}
\caption{The main screen a student views after logging on to the $R2R$ system using a smartphone.}
\label{fig:session}
\vspace{0.6cm}
\end{figure}

\begin{figure}[h]  
\centering
\includegraphics[scale = 0.55]{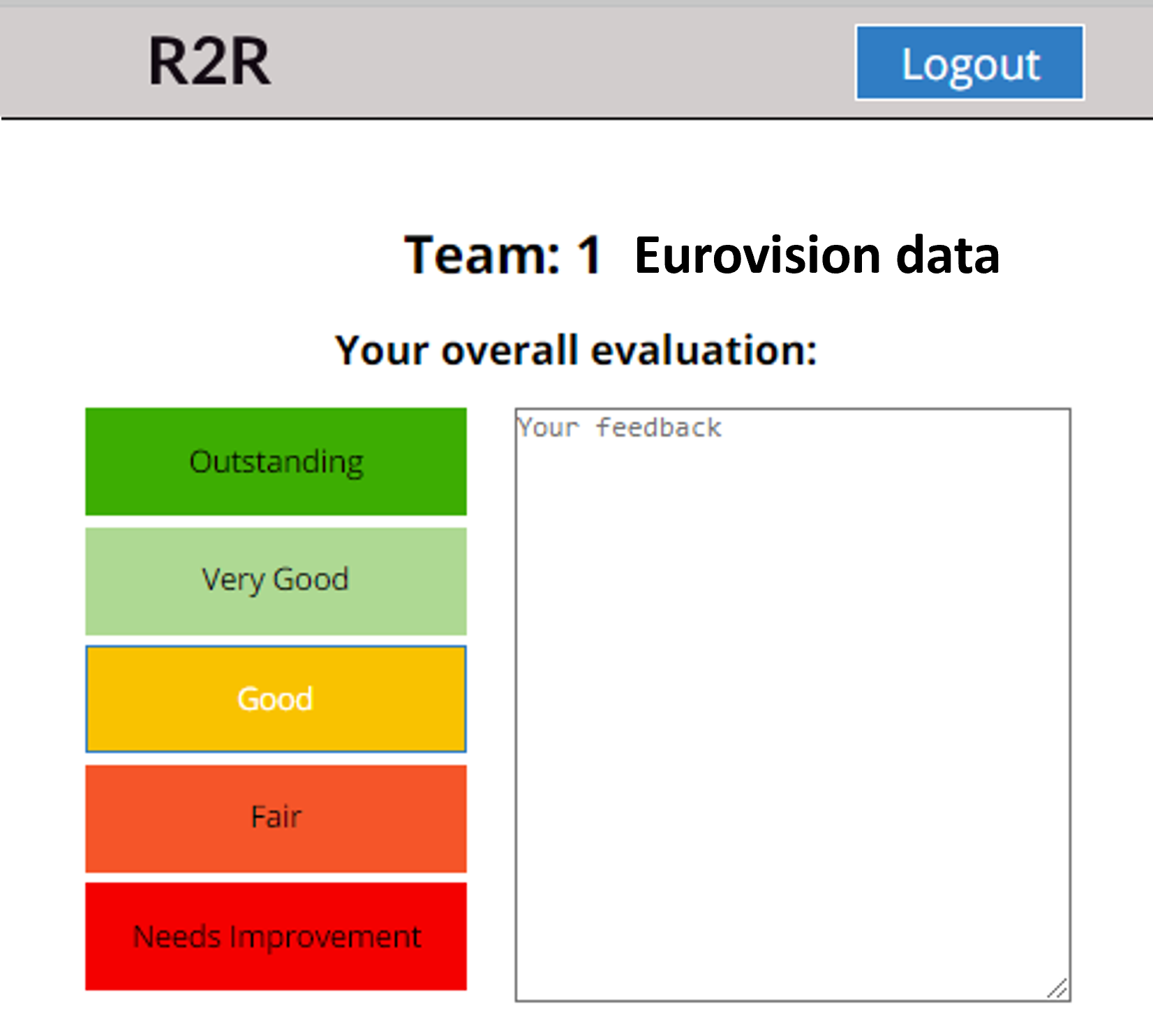}
\caption{The evaluation screen of a specific project. }
\label{fig:question}
\vspace{1cm}
\end{figure}

\begin{figure}[h] 
\centering
\includegraphics[scale = 0.2]{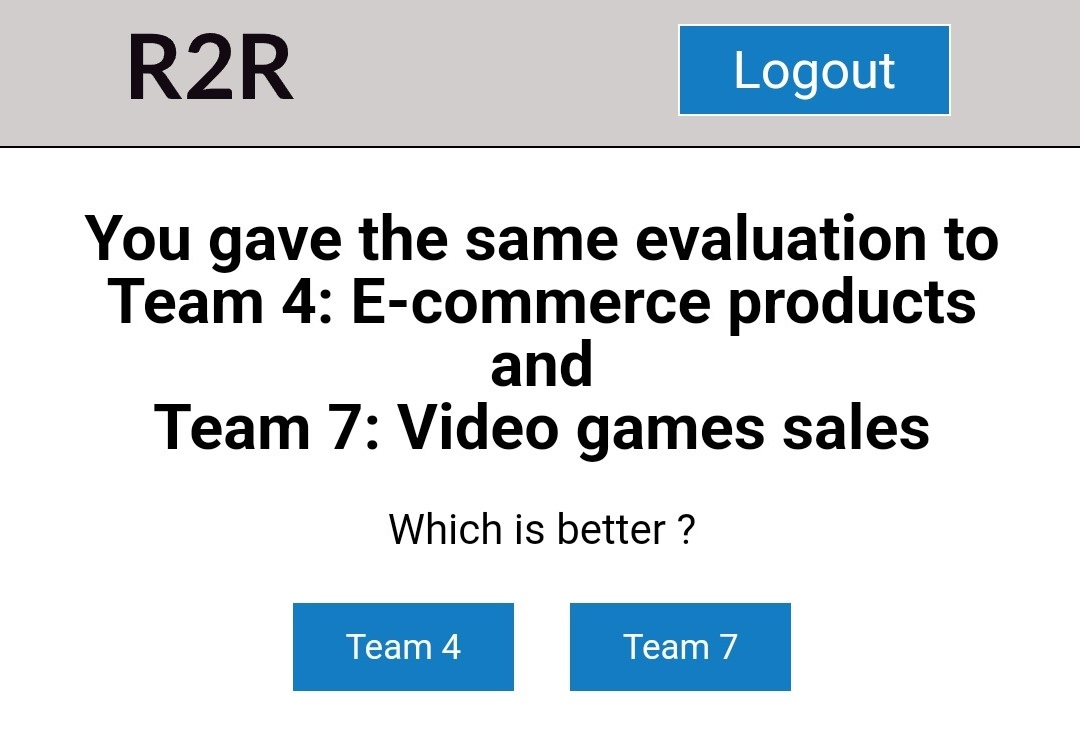}
\caption{$R2R$ Pairwise comparison question displayed to a student after they assigned the same evaluation to two projects. }
\label{fig:pairwise}
\vspace{0.5cm}
\end{figure}

To initialize the system, project names and numbers were fed into the system beforehand by a system admin. A link to the system was then distributed to the students.
Figures \ref{fig:session}-\ref{fig:pairwise} present screenshots from the $R2R$ system interface as seen on a smartphone. 
Students are met with the main screen after logging into the $R2R$ system (Figure \ref{fig:session}). This screen displays project numbers and names. Grey boxes indicate that evaluations for those projects have been successfully completed, while blue boxes signal that evaluations are still pending and can be submitted. Additional projects become visible as students scroll down the screen.  

After each presentation, students were asked to use the system to assess the quality of the project. Upon selecting a box to initiate input, a project-specific evaluation screen emerges. Within this interface, students are presented with a set of five radio buttons, each corresponding to different evaluation levels. 
We adopted a grading scale consisting of five grades ($D = {1, 2 \ldots 5 }$), based on the research of~\citet{miller1956magical}, which suggested that people can distinguish between seven plus or minus two levels of intensity. To minimize bias in the ratings, we utilized a standardized grading language proposed by ~\citet{balinski2011majority} that maps to the following five grades: Outstanding (5), Very Good (4), Good (3), Fair (2), and Needs improvement (1). 
Students were instructed to evaluate each project holistically and in accordance with the specified project requirements. A project graded as ``Outstanding'' met the requirements in an exceptional manner. 

While an option for providing written feedback is available, it remained non-mandatory (Figure \ref{fig:question}).
In the event that a student assigned the same grade to more than one project, pairwise comparison queries were executed following algorithm \ref{alg:R2R} (Figure \ref{fig:pairwise}). 
Finally, the system produced ranked rating sets, one set for each participating student.

\subsection{Data collection}\label{subsec:data}
The system was deployed as part of a three-credit course on Data Analysis at Ariel University. 
Students presented their final projects in front of the class as part of the course. 
The other students present were instructed to input their evaluations into the $R2R$ system following each project presentation.

The full data collection statistics are presented in Table \ref{table:stats}.
Each student participated in one class presentation session. In each session, 8-12 projects were evaluated. We held 10 different sessions. The study received ethical approval from the institutional review board of the university. To maintain the participants' privacy, data was anonymized during the analysis, storage, and reporting stage, i.e., session dates, student identification numbers, and project names were concealed. 

In total, 288 students participated in the experiments.
We removed 13 students who did not rate all the projects in their session. 
Another 31 students did not report their preferences iteratively, according to the project presentation order. These students were also removed, as they might have reported arbitrary preferences.  
Thus, the responses of a total of 244 students were included in the data analysis.

% Please add the following required packages to your document preamble:
% \usepackage{multirow}
\begin{table}[h!]
\caption{Data collection statistics.}
\label{table:stats}
\begin{tabular}{@{}ccccl@{}}
\toprule
\multirow{3}{*}{Session} & 
\multirow{3}{*}{
    \begin{tabular}[c]{@{}l@{}}
    Number of\\ participants\end{tabular}} & 
\multirow{3}{*}{
\begin{tabular}[c]{@{}l@{}}
    Number of\\ projects\end{tabular}} & 
\multirow{3}{*}{
    \begin{tabular}[c]{@{}l@{}}
    Number of median grades\\ (Outstanding, Very Good, \\ Good, Fair, Needs improvement)
\end{tabular}} &  \\
 &  &  &  &  \\
 &  &  &  &  \\ 
\cmidrule(r){1-4}
\Romannum{1}  & 27 & 10 & 6,4,0,0,0    \\ 
\Romannum{2}  & 24 & 10 & 5,4,1,0,0   \\  
\Romannum{3}  & 29 & 10 & 4,6,0,0,0  \\  
\Romannum{4}  & 22  & 9 & 2,7,0,0,0   \\  
\Romannum{5}  & 27  & 10 & 3,5,2,0,0   \\  
\Romannum{6}  & 34  & 12 & 0,8,4,0,0   \\ 
\Romannum{7}  & 18  & 8 & 1,4,3,0,0   \\
\Romannum{8}  & 21  & 9 & 3,6,0,0,0   \\
\Romannum{9}  & 20  & 9 & 4,5,0,0,0   \\
\Romannum{10}  & 22  & 10 & 2,6,2,0,0   \\
\bottomrule
\end{tabular}
%\vspace{0.3cm}
\end{table}

Most projects received a median grade of either ``Outstanding'' or ``Very Good''. For example, in Session \Romannum{1} (the first row in Table \ref{table:stats}), 27 students were asked to rate ten projects. Six projects received the median grade of ``Outstanding'', four received the median grade of ``Very Good''. %and no projects received the median grade of ``good''. 
All data and the Python notebook code written for the evaluation are available upon request.

\section{Evaluation}
The evaluation focused on two key aspects: 
\begin{itemize}
    \item \textbf{Methodological bias:} The project presentation order is predetermined. Students evaluate the projects iteratively after each presentation. While this is inherent to the problem setting and cannot be altered, it may influence student evaluations and, consequently, the final project rankings. We examined two potential biases: one related to grading and the other to comparing. In Section~\ref{subsec:bias}, we analyzed the presentation order bias, where students might assign higher (or lower) grades to the projects presented at the beginning. In Section~\ref{subsec:primacy-recency}, we investigated primacy and recency bias, wherein students, when asked to choose between two projects to which they assigned the same grade, may preferentially select the first or latter project.
    \item \textbf{R2R effectiveness:} We compared the effectiveness of $R2R$ to other peer assessment methods. In Section~\ref{subsec:comm}, we quantified the reduction in communication load when using $R2R$ compared to methods requiring pairwise comparisons. In Section~\ref{subsec:ties}, we measured the extent to which $R2R$ reduces ties compared to other median-based voting rules and to the computation of the average. In Section~\ref{subsec:impact}, we compared the rankings produced by $R2R$ with those derived from other methods. 
\end{itemize}

\subsection{Presentation order bias}\label{subsec:bias}
\begin{figure}[h] 
\centering
\includegraphics[scale=0.4]{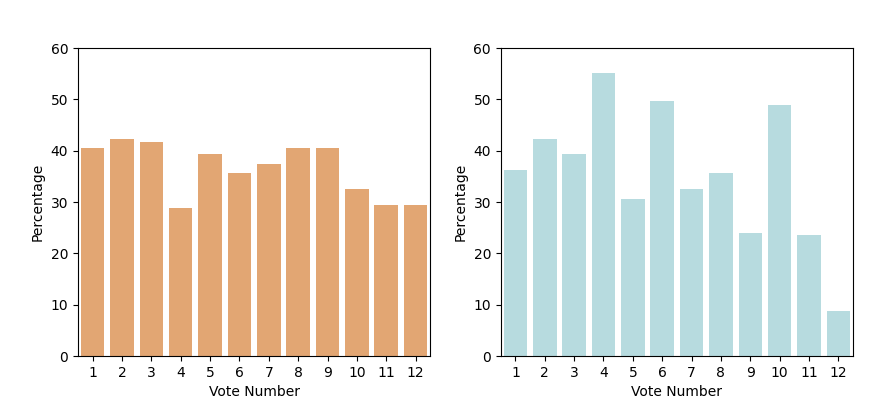}
\caption{Percentage of voters that assigned a score of ``Outstanding'' (left) and ``Very Good'' (right).}
\label{fig:voting_perc}
%\vspace{0.5cm}
\end{figure}

We explored the potential influence of presentation bias on project grading. 
We hypothesized that students might assign higher grades to the first projects presented. Subsequently, as more projects are presented and students engage in pairwise comparison prompts, they may distribute their scores more evenly, possibly as a strategy to circumvent the prompts. 
However, Figure \ref{fig:voting_perc}
dispelled these notions. The y-axis is the percentage of voters that assigned a score of ``Outstanding'' (left figure) and ``Very Good'' (right figure). 
On the x-axis, projects are arranged according to their grading order, from the first project (numbered as project 1) to the last (either project 8, 9, 10, or 12, depending on the session). 
%The absence of leftward or rightward skewness in the plots indicates no observable trend of initiating with high scores and subsequently reducing or increasing them.
In the sole session featuring 12 projects, ``Very Good'' was assigned by few to the 12th project. However, this outcome may be attributed to the project's inherent qualities. 
%Note: projects were NOT presented according to their number. i.e., project #12 may have been presented first
%ADDED 19.6
We randomly shuffled project orders 100 times to compute an average ``Outstanding'' score distribution. Comparing it with the distribution in~\ref{fig:voting_perc}, a chi-square test could not reject the null hypothesis that both distributions are the same($p = 0.23$). We repeated this for ``Very Good'' scores, yielding similar results  ($p = 0.26$). 

\subsection{Primacy and recency bias}\label{subsec:primacy-recency}
Next, we examined whether users are prone to primacy or recency bias during pairwise comparisons. In other words, when prompted with a pairwise query, do users tend to select the first or last projects they see? 
To evaluate this, we analyzed instances where projects received the same score within a session. Subsequently, we computed the percentage of cases where projects were ranked in ascending or descending order. For example, if a student saw projects in this order of presentations: $c_1,c_2 \ldots c_{10}$, and rated projects $c_2, c_3, c_8$ as``Outstanding'', and after answering pairwise comparisons, their preference profile is revealed as: $c_2 \succ c_3 \succ  c_8$, it suggests a potential preference for the first projects seen --- a primacy bias.  
Conversely, if a student rated projects $c_1$ and $c_5$ as ``Very Good'' and preferred $c_5$ to $c_1$, it suggested a preference for the last projects seen --- a recency bias.
Importantly, students might display a recency bias in some evaluations and a primacy bias in others. The results are reported in Table \ref{table:bias}. 
%Added 19.6
In a simulation study, we randomly shuffled the ordering of the projects that received a similar grade from each student. This process was iterated 100 times, from which the average primacy and recency bias proportions were derived. Subsequently, we utilized a two-sided t-test to compare the simulated primacy proportions with those presented in Table 2.  The test could not reject the null hypothesis that the proportion means are the same ($p =  0.39$). The same test was conducted for the recency bias, yielding similar results ($p = 0.91$).

%, with no discernible trend observed towards primacy or recency. 
%The data for ``Needs improvement'' ratings is insufficient to discern a trend, as this grade was assigned only 42 times (1.7\% of all grades).  

\begin{table}
\centering
\caption{Primacy and recency bias proportions.}
\begin{tabular}{@{}lll@{}}
 \toprule
  Evaluation  & 
  Primacy bias  & 
  Recency bias  \\
 \midrule
  Outstanding  & 0.19 & 0.16     \\
  Very Good  & 0.17 & 0.1     \\
  Good  & 0.19 & 0.3     \\
 Fair  & 0.32  & 0.42     \\
 Needs improvement  & 0.14  & 0.43     \\
\bottomrule
\end{tabular}
\label{table:bias}
\end{table}

\subsection{Communication load}\label{subsec:comm}
Table \ref{table:communication} summarizes the communication load findings.
For 8,9,10, and 12 projects, the maximum pairwise queries are 28,36,45, and 66, respectively. In the ten sessions, the average number of issued pairwise queries lay in the range of 6.5-14.6 (the standard deviation is displayed in parentheses).  
Thus, $R2R$ reduces the communication load by $62\% - 77\% $. In other words, on average, students had to respond to less than $30\%$ of the total number of possible pairwise queries. 

\begin{table}[h!]
\caption{Communication load statistics}
\centering
\begin{tabular}{@{}lll@{}}
\toprule
\multirow{2}{*}{Session} & 
\multirow{2}{*}{\begin{tabular}[c]{@{}l@{}}Pairwise comparison\\ queries\end{tabular}} & 
\multirow{2}{*}{\begin{tabular}[c]{@{}l@{}}Communication\\ reduction\end{tabular}} \\
 &  &  \\ 
 \midrule
 \Romannum{1}  & 13.9(5.7) & 69\%   \\
 \Romannum{2}  & 10.8(2.7) & 75\%  \\
 \Romannum{3}  & 13.2(5.0) & 71\%  \\
 \Romannum{4}  & 11.8(3.9)  & 67\% \\
 \Romannum{5}  & 10.6(2.5)  & 75\%  \\
 \Romannum{6}  & 14.6(5.9)  & 77\%  \\
 \Romannum{7}  & 6.5(2.5)  & 76\% \\
 \Romannum{8}  & 10.4(2.5)  & 71\% \\
 \Romannum{9}  & 13.5(6.8)  & 62\% \\
 \Romannum{10}  & 10.6(3.1)  & 76\% \\
\bottomrule
\end{tabular}
\label{table:communication}
%\vspace{0.3cm}
\end{table}

\begin{table*} [ht]
\caption{Average (and standard deviation) of Kendall distance between different methods. }
\centering
\begin{tabular}{@{}lllllll@{}}
\toprule
\multirow{2}{*}{} & 
\multirow{2}{*}{R2R\_b} & 
\multirow{2}{*}{R2R\_c} &
\multirow{2}{*}{
    \begin{tabular}[c]{@{}l@{}}
    Typical\\ ranking
    \end{tabular}} & 
\multirow{2}{*}{
    \begin{tabular}[c]{@{}l@{}}
    Central\\ ranking
    \end{tabular}} & 
\multirow{2}{*}{
    \begin{tabular}[c]{@{}l@{}}
    Usual\\ ranking
    \end{tabular}} & 
\multirow{2}{*}{
    \begin{tabular}[c]{@{}l@{}}
    Majority\\ ranking
    \end{tabular}}  
\\
 &  &  \\ 
\midrule
 R2R\_c &           1.3 (0.4)& & & & & \\
 Typical ranking  & 2.3 (2.1) & 2.9 (1.5) &           &           &           & \\
 Central  ranking & 3.4 (1.4) & 4.0 (1.3) & 2.4 (1.3) &           &           & \\
 Usual  ranking   & 2.4 (2.1) & 2.9 (1.7) & 0.6 (0.5) & 3.1 (1.1) &           & \\
 Majority ranking & 2.8 (2.0) & 2.9 (2.4) & 1.3 (0.9) & 3.9 (1.5) & 0.8 (0.7) & \\
 Mean &             2.6 (2.1) & 3.2 (1.4) & 0.6 (0.4) & 2.4 (1.2) & 1.1 (0.7) & 1.6 (1.1)   \\
\bottomrule
\end{tabular}
\label{table:kendall}
\end{table*}

\subsection{Ties in rankings}\label{subsec:ties}
We studied two instances of the proposed $R2R$ method: $R2R\_b$ and $R2R\_c$. They use Borda and Copland voting, respectively, in the aggregation phase (see Section~\ref{subsec:agg}). These instances were compared to existing methods: {\sc Majority ranking}~\citep{balinski2011majority}, {\sc Typical ranking}, {\sc Central ranking}, and {\sc Usual ranking}. The three latter are the rankings retrieved from the Typical judgment, Central judgment, and Usual judgment methods~\citep{fabre2021tie} (see Section~\ref{sec:lit}). We also compared these methods to a simple {\sc Mean} method that computed the average rating of each alternative and ranked the alternatives accordingly. 

Experiments were conducted with a varying number of voters (students in the user study), ranging from 2 to 20 (18 for Session \Romannum{7}), which were sampled without replacement. Each experiment ran $50$ times. 
The $\alpha$ in $Copeland^{\alpha}$ was set to $\alpha = \frac{1}{3}$, as it provided better results than $\alpha = \{0, \frac{1}{2}, 1\}$
(setting $\alpha$ to values smaller than $\frac{1}{3}$ did not yield an improvement). 

\begin{figure} [ht]
\centering
\includegraphics[width=3.2in]{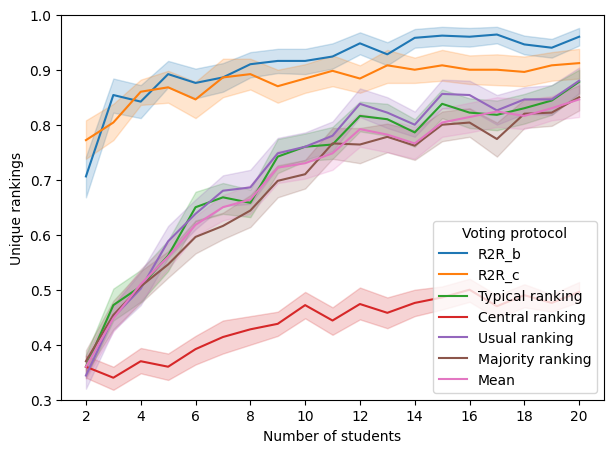}
\caption{ Ties in rankings in Session \Romannum{1}.}
\label{fig:a_unq}
\vspace{0.8cm}
\end{figure}

The unique number of rankings in each method was computed, which refers to the number of projects not tied in their order.
Figure \ref{fig:a_unq} compares all methods in Session \Romannum{1}. 
Axis y is the fraction of unique rankings (with 1 meaning there are no ties in the final ranking). 
All sessions display a similar trend; as the number of voters increased, the methods exhibited better performance with fewer ties and more unique rankings. Note that even the {\sc Mean} method has ties in rankings. 
Employing $R2R\_b$ or $R2R\_c$ results in the fewest ties in the final ranking.
In all sessions, $R2R\_b$ exhibited better performance than the other methods,  especially when the number of voters was small. 
For an odd number of voters, $R2R\_c$ outperforms the other methods in some sessions. 
This advantage stems from the fact underlying the Copeland voting rule. Consequently, the observed increase in performance may not exhibit a smooth progression.
Comparisons for the rest of the sessions demonstrate similar results (Appendix \footnote{\url{https://doi.org/10.17605/OSF.IO/K4YMA}}).

To statistically compare the seven methods, 
we followed the robust non-parametric procedure described in \citep{garcia2010advanced}.
%\citep{demvsar2006statistical, garcia2010advanced}.
The Friedman Aligned Ranks test with a confidence level of 95\% rejected the null hypothesis that all seven methods perform the same.
Further applying the Conover post hoc test supported our findings; 
$R2R\_b$ significantly outperformed all methods (except $R2R\_c$) with a confidence level of 95\%. 
$R2R\_c$ significantly outperformed all methods but {\sc Usual ranking} with a confidence level of 95\%. 
No statistical difference was found between $R2R\_b$ and $R2R\_c$.

\subsection{Consistency and tie-breaking impact}\label{subsec:impact}
In line with the findings of \citet{fabre2021tie}, our investigation focused on assessing the degree to which different methods yield consistent outcomes while considering the impact of tie-breaking on these results. To quantify this impact, we calculated the Kendall-tau %\citep{kendall1948rank}
distance between the various rankings. The computed distances are presented in Table \ref{table:kendall}. A Kendall distance of 5 indicates that, on average, there is a 5\% likelihood that the relative order of two items will be reversed.

Our results align with those of \citet{fabre2021tie}, showing that in more than 95\% of instances, the methods yield equivalent results. For the remaining cases, tie-breaking rules determine the outcome. %Notably, the method $R2R\_c$, based on the Copeland voting rule, consistently demonstrates higher agreement with the other methods in comparison to $R2R\_b$. 
 
\section{Discussion}
$R2R$ originated from one of the authors' needs for a method to conduct peer assessment during oral presentations in the classroom, driven by the inadequacy of existing options. It provides a hybrid approach combining rating-based evaluations with pairwise comparison queries when necessary.

$R2R$ offers a structured methodology for eliciting voter preferences through cardinal evaluations, such as scores or grades, while also accommodating ordinal preferences through pairwise comparisons. By aggregating these preferences, the $R2R$ method generates a ranked list of alternatives that reflects students' collective opinions. 

While the $R2R$ model may be considered naive in its structure and implementation, its effectiveness stems from its ease of understanding, use, and minimal assumptions, echoing the principles of Occam's razor. This inherent simplicity enhances its potential for widespread adoption and adaptation.

In a user study involving nearly 300 students, we did not identify any methodological bias in presentation order or primacy-recency bias. Furthermore, the effectiveness of the $R2R$ became evident through several key findings. 
Firstly, we found that $R2R$ outperformed existing methods by reducing the number of ties in the ranked output, particularly notable for small groups of voters, as often happens in in-class presentations. 
Moreover, $R2R$ imposes a lower communication load on students compared to pairwise comparison methods, as students only need to evaluate each alternative once instead of repeatedly comparing pairs of alternatives. The cognitive load is also diminished compared to models that necessitate students to rank all alternatives, as students only need to assign a score or grade to each alternative. 

One of the by-products of the user study is the acquisition of both project ratings and rankings, resulting in a unique dataset that may be utilized for further studies.

We presented two variations of the $R2R$ model, each employing a distinct tie-breaking method. The first variation utilizes the Borda voting rule for tie-breaking ($R2R\_b$), while the second variation employs the Copeland voting rule ($R2R\_c$). $R2R\_b$ excelled in its tie-breaking capabilities and exhibited greater consistency with the outcomes of all other methods. Thus, we recommend its use. It is also feasible to utilize other voting rules. 

% % Young 1986 paper: Condorcet said that when there are two alternatives, majority voting is best. When there are more than two alternatives, find the candidate that beats the rest. Young says Borda is best for choosing the best candidate, and Condorcet for choosing the best ranking. 

Although the $R2R$ system was initially designed and implemented for a classroom setting, its potential applications extend beyond academia. In the workplace, for instance, group or team leaders often need to evaluate the performance of their employees, as noted by \citep{arvey1998performance}. This is a sensitive task since an employee's position on the final performance list usually determines the bonus they will receive. Managers can use the $R2R$ to establish their personal ranked list of employees while minimizing communication and cognitive overload.

\textbf{Limitations.} 
The current implementation used a 5-point Likert scale.
Employing a 7-point or 9-point scale could reduce communication load (fewer pairwise queries) but might increase cognitive load and bias (more options to choose from). This aspect remains underexplored.
Additionally, the model assumes students will respond truthfully to pairwise comparison queries. However, the possibility of strategic behavior remains, as discussed in ~\citep{dery_lieOnTheFly,dery2019lie}, and its impact on the results is not fully explored in this paper.

 % It's also worth noting that while the primacy-recency bias has been studied in the context of memory~\citep{murphy2006primacy} and decisions such as hyperlink selection~\citep{lee2022exploring}, we did not find research on primacy-recency effects in peer assessment. This area is worth further exploration and an extended evaluation.

\textbf{Future work.} In the current model, students cannot express indifference between projects. While this approach ensures individual tie-free rankings, it might inadvertently restrict the representation of students' true preferences. Future investigations should consider the implications of relaxing this restriction and explore how it impacts the quality of the individual output rankings and the aggregation of the rankings. Another avenue is to examine the incorporation of multiple pairwise comparison types, as suggested by \citet{newman2022ranking}. %Lastly, we did not find research on primacy-recency effects in peer assessment. This area is worth further exploration and an extended evaluation.

%Overall, $R2R$ provides a simple and effective approach for peer assessment and presents a robust and versatile approach to the challenge of converting ratings into meaningful rankings. Its demonstrated advantages in tie reduction, reduced communication load, and applicability in diverse contexts highlight its potential contributions to decision-making and preference aggregation, both within academic contexts and beyond. 

%add to futute work?
% \textbf{the impact of these methodologies on the assessors and the assessment}. 
%Another potential direction is to integrate a ``truth serum", as recommended by \citep{prelec2004bayesian}, which would incentivize voters to provide their honest preferences as they would be evaluated. These research directions could further enhance the usefulness and applicability of the $R2R$ system in various domains.

%%%%%%%%%%%%%%%%%%%%%%%%%%%%%%%%%%%%%%%%%%%%%%%%%%%%%%%%%%%%%%%%%%%%%%%%

%%% Use this environment to include acknowledgements (optional).
%%% This will be omitted in doubleblind mode.

% \begin{ack}
% We want to extend our special thanks to Matan Lange for implementing the $R2R$ system and for his valuable assistance during the user study. We also appreciate Koby Karady's aid in the data analysis of order bias. 
% \end{ack}

%%%%%%%%%%%%%%%%%%%%%%%%%%%%%%%%%%%%%%%%%%%%%%%%%%%%%%%%%%%%%%%%%%%%%%%%

%%% Use this command to include your bibliography file.

%\bibliography{refs}

\end{document}